\input{aipcheck}

\documentclass[
    ,final            
  ]
  {aipproc}

\layoutstyle{6x9}

\begin{document}

\title{Pulsar Prospects for the Cherenkov Telescope Array}


\author{T. Hassan}{
  address={Universidad Complutense de Madrid, Spain}
}

\author{S. Bonnefoy}{
  address={Universidad Complutense de Madrid, Spain}
}

\author{M. L\'{o}pez}{
  address={Universidad Complutense de Madrid, Spain}
}

\author{N. Mirabal}{
  address={Universidad Complutense de Madrid, Spain}
}

\author{J. A. Barrio}{
  address={Universidad Complutense de Madrid, Spain}
}

\author{J. L. Contreras}{
  address={Universidad Complutense de Madrid, Spain}
}

\author{R. de los Reyes}{
  address={Max-Planck-Institut f\"{u}r Kernphysik, Heidelberg, Germany}
}

\author{E. O. Wilhelmi}{
  address={Max-Planck-Institut f\"{u}r Kernphysik, Heidelberg, Germany}
}

\author{B. Rudak}{
  address={Nicolaus Copernicus Astronomical Center, Toru\'{n}, Poland}
}

\author{for the CTA Consortium}{
  address={See \url{http://www.cta-observatory.org} for full author \&
affiliation list}
}

\begin{abstract}
In the last few years, the {\it Fermi}-LAT telescope has discovered 
over a 100 pulsars at energies
above 100 MeV, increasing the number of known gamma-ray pulsars by an order
of magnitude. In parallel, imaging Cherenkov telescopes, such as MAGIC and
VERITAS, have detected for the first time VHE pulsed gamma-rays from the
Crab pulsar. Such detections have revealed that the Crab VHE spectrum follows a
power-law up to at least 400 GeV, challenging most theoretical models, and
opening wide possibilities of detecting more pulsars from the ground with 
the future
Cherenkov Telescope Array (CTA). In this contribution, we study 
the capabilities
of CTA for detecting {\it Fermi} pulsars. For this, we extrapolate 
their spectra with
"Crab-like" power-law tails in the VHE range, as suggested by the latest MAGIC
and VERITAS results.
\end{abstract}

\maketitle


\section{INTRODUCTION}

  The detection of neutron stars through gamma-ray
pulsations is a key science goal for the future Cherenkov
Telescope Array (CTA) \citep{CTA}. Gamma-ray pulsar observations at
high energies (over a few tens of GeV) could help to
understand the region where pulsed emission takes place
by comparing the measured spectra with predictions by
theoretical models.
The {\it Fermi} mission has revolutionized the study of gamma-ray
pulsars detecting 117 sources in the MeV-GeV
energy range, which are reasonably fitted with sharp
exponential cutoff values between 0.7 to 7.7 GeV \citep{abdo}.

Nonetheless, the detection of the Crab pulsar above 25 GeV with IACTs
\citep{aliua,aliub,aleksic} has reframed the 
exponential cutoff observed by {\it Fermi}
 in favor of a broken power-law shape
that extends the pulsed emission up to 400 GeV. This 
recent discovery motivates the need for further
pulsar studies in the VHE regime.

\begin{figure}
\hfil
\includegraphics[width=16cm, height=10cm]{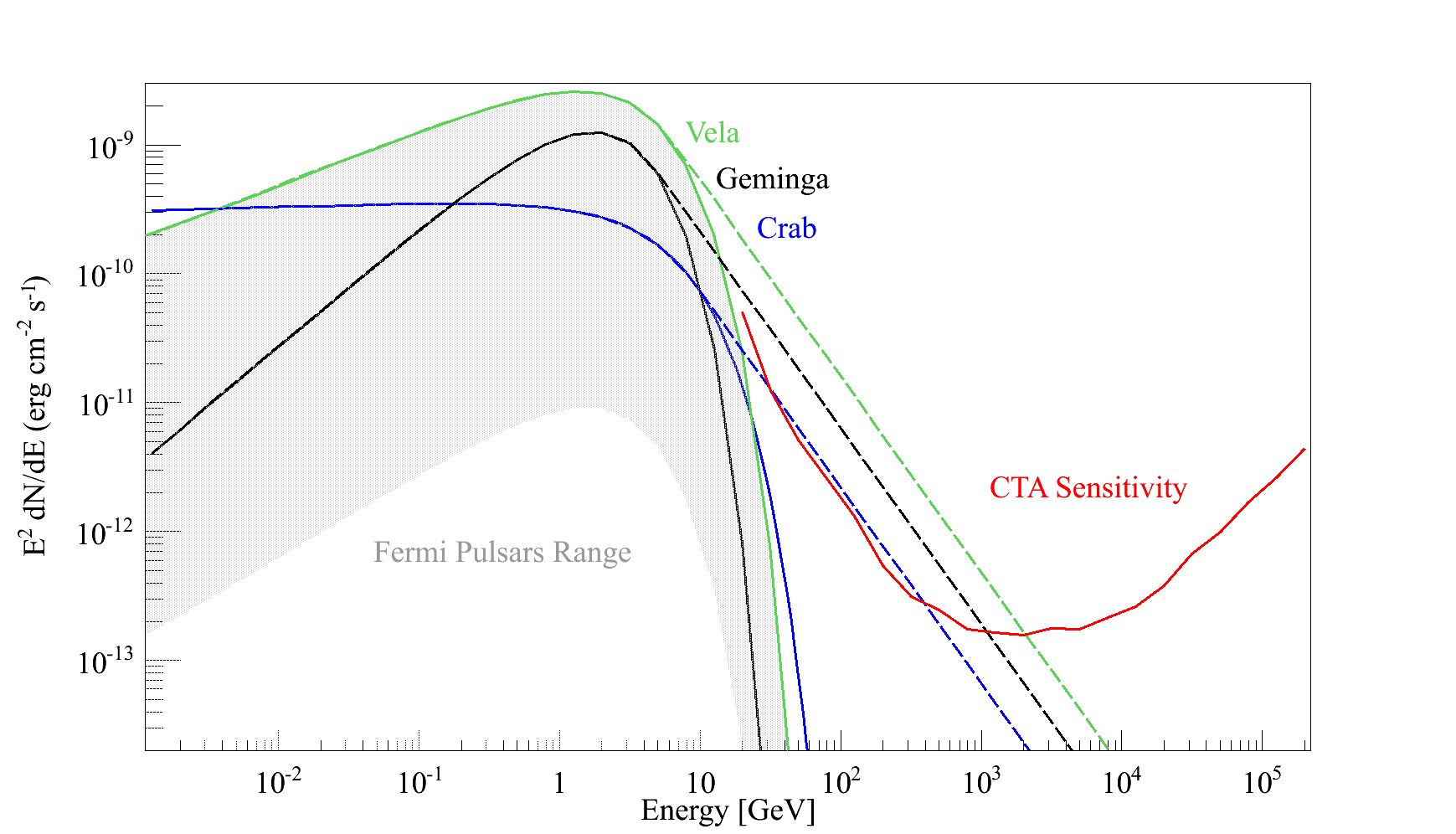}
\hfil
\caption{Fermi-LAT pulsars general profile (grey area) with standard CTA
sensitivity curve for configuration B in 50h. Vela, Geminga and Crab
extrapolated SEDs (dashed lines).}
\label{flux_index}
\end{figure}

To place this in context, Fig. \ref{flux_index} shows 
the spectral fits (power-law with exponential cutoff)
for 46 {\it Fermi} pulsars taken from \citep{abdo}, 
in comparison with the standard CTA diferential sensitivity
curve for configuration B (configuration with 5 LSTs \cite{CTA}) in 50 h . The fits of Vela, Crab  and
Geminga pulsars are indicated explicitly, while the shaded area contains 
the fits for the remaining 43 pulsars.

In order to estimate Cherenkov Telescope Array (CTA) potential, 
we 
generate a full simulated spectrum for a 50 h observation of the Crab pulsar
\citep{emma}. 
We also explore the detectability of Fermi pulsars presuming a
"Crab-like" power-law tail, as the latest observations
suggest.

\section{PROCEDURE}

Initially, we generate a 50 h simulated observation of the 
Crab pulsar using \textit{CTAmacrosv6}.
Total emission (P1 + P2) and both P1 and P2 peaks were simulated 
using the MAGIC power-law fits given in \cite{aleksic}. CTA 
configuration B, E and C \cite{CTA} were tested.

To explore the detectability of \textit{Fermi} pulsars, we extend their spectra above 
the cutoff energy with a power-law tail that
assumes the same spectral index ($\beta$) as the one 
found for the Crab, when
a broken power-law is applied to fit both {\it Fermi} LAT 
and VERITAS detections, i.e
$\beta = 3.52$ \citep{aliub}. In Fig. \ref{flux_index} we show some examples of the extrapolation performed in dashed lines. 

We consider a 90\% background reduction
assuming a pulsed duty cycle of 10\% (reduced to 5\% for the 
Crab pulsar independent peaks), systematic errors of 5\% and standard 
detection conditions (detection over $5\sigma$ in 50 
hours of observation time). 
No gamma-ray emission from a pulsar wind nebula was considered.

\section{RESULTS}

\begin{figure}
\hfil
\includegraphics[width=16cm, height=9.5cm]{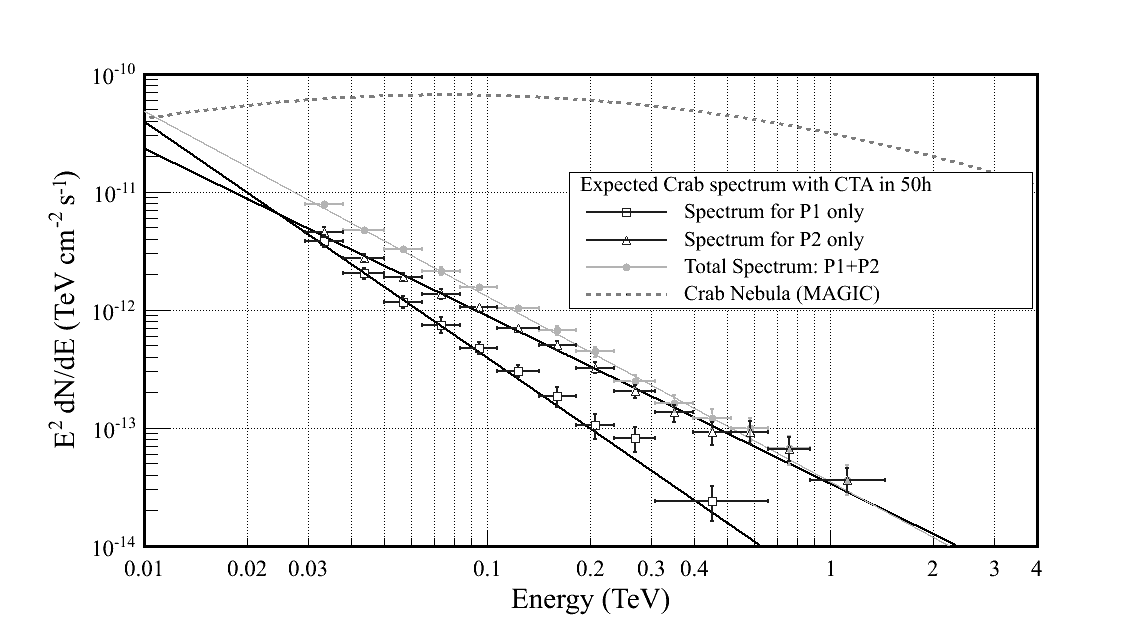}
\hfil
\caption{Simulated spectra of each of the two Crab phase peaks for 50 h with CTA configuration B and total spectrum 
using MAGIC power-law fits given in \cite{aleksic}.}
\label{Crab_pulsar}
\end{figure}

In Fig. \ref{Crab_pulsar} we show the simulated 50 h observation of 
CTA array B of the Crab pulsar. One can see how both total signal 
and resolved peaks are well characterized up to 1 TeV. 

\begin{figure}
\hfil
\includegraphics[width=15cm, height=9cm]{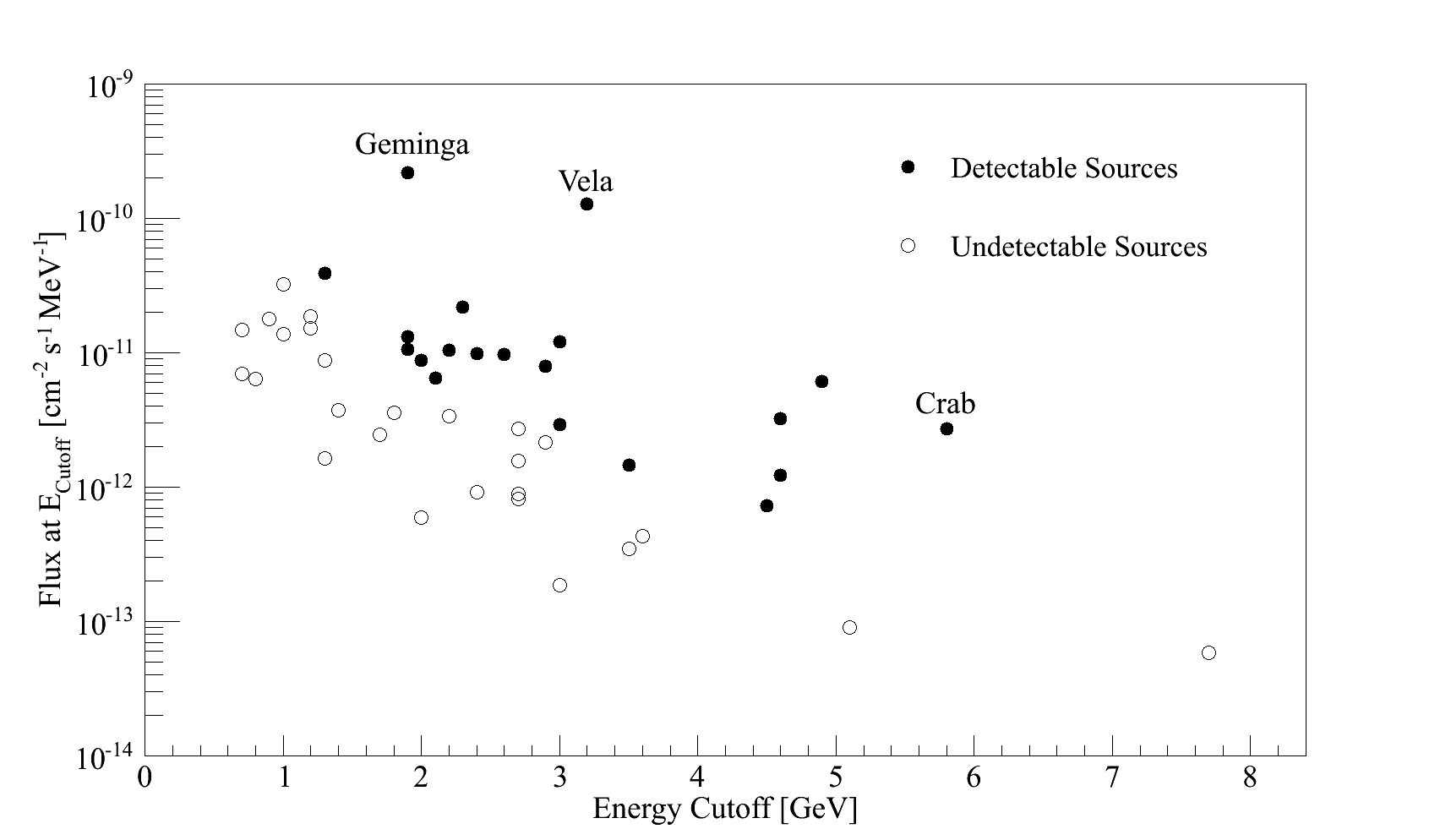}
\hfil
\caption{Detectable pulsars extrapolated with Crab pulsar power-law index
($\Gamma$=3.57, from [3]) by the future CTA array B in 50h.}
\label{detectable}
\end{figure}

Presuming the previous hypothesis of ``Crab-like`` emission, we found 
that 20 pulsars
would be detectable by CTA with configurations B and E. 
Figure \ref{detectable} shows how the detectability with configuration
B depends on the exponential cutoff energy value (as determined by
 {\it Fermi} LAT)
and the photon flux density at this energy.

\section{CONCLUSIONS}

CTA potential for pulsar detection seems encouraging, as it will be able to 
reveal the extent of the Crab pulsed emission up to at least 1 TeV. 
In fact, the bare detection of the pulsations would take less than one hour.

Under the hypothesis that the existence of VHE-tails is a universal feature 
in pulsars, we
conclude that 20 pulsars would be detectable by CTA. This represents a 
large fraction (up to
40\%) of the brightest {\it Fermi} pulsars. We can also affirm 
that CTA configurations with more Large Size 
Telescopes (LST) are preferred, as pulsar detection falls 
in the low energy range (50 to 200 GeV). 

Needless to say, there is no assurance that gamma-ray pulsars 
will cooperate in the way described above.
However, some theoretical models of young and energetic 
pulsars as well as old
millisecond pulsars speak in favor of pulsed spectral 
components located in the
VHE domain \citep{aleksic2011}. CTA will be the only facility in 
near future capable of solving this problem.


\begin{theacknowledgments}
  The authors acknowledge the support of the Spanish MINECO under
project code FPA2010-22056-C06-06. M.L. and N.M. gratefully
acknowledge support from the Spanish MINECO through a Ram\'{o}n y Cajal
fellowship. The authors also gratefully acknowledge support from the
agencies and organizations listed in this page: \url{http://www.cta-observatory.org/?q=node/22}.
\end{theacknowledgments}





\end{document}